\documentclass{aa}

\usepackage{graphicx}

\usepackage{txfonts}

\usepackage[labelfont=bf,justification=justified]{caption}
\DeclareCaptionFormat{cont}{#1\textbf{.continued}#2#3\par}
\usepackage{indentfirst}
\usepackage[title]{appendix}
\usepackage[usenames]{color}
\usepackage[export]{adjustbox}
\usepackage{float}
\usepackage{hyperref}

\newcommand{\logg}{$\log g$}
\newcommand{\filou}{{\sc{filou}}}
\newcommand{\vsini}{$\mathrm{v}\cdot\!\sin\!~i$}
\newcommand{\kms}{$\mathrm{km}~\mathrm{s}^{-1}$}

\newcommand{\multim}{{\sc MultiModes}}
\newcommand{\mesa}{{\sc mesa}}
\newcommand{\gyre}{{\sc gyre}}
\newcommand{\myr}{$\mathrm{Myr}$}

\begin{document}

   \title{Asteroseismology of 35 Kepler and TESS $\delta$ Scuti stars near the red edge of the instability strip}

   \subtitle{The limitations of $\delta$ Scuti stars for dating open clusters}

   \author{D. Pamos Ortega}

   \institute{Departamento de Física y Química. IES Levante. Avenida Italia s/n. 11205. Algeciras. Spain\\
              \email{david.pamos.or@gmail.com}
             }
             
   \date{Received 25 July 2025 / Accepted 3 September 2025}

  \abstract
   {}
   {The aim of this work is to determine the maximum ages that can be unambiguously established for $\delta$ Sct stars using seismic observables, and, by extension, the oldest open clusters that can be dated using this type of star.}
   {I estimate the large frequency separation using various techniques applied to two samples of $\delta$ Sct located near the red edge of the instability strip. One sample consists of 18 targets observed by the Kepler mission, and the other comprises 17 targets observed by TESS. I employ a grid of stellar models representative of typical $\delta$ Sct parameters, incorporating mass, metallicity, and rotation as independent variables, and compute the first eight radial modes for each model.  
   Using the observed spectroscopic temperature, and the estimated large separation, I construct mode identification boxes by matching observed frequencies with those predicted by the models. These matches allow for the identification of possible radial modes. Based on this information, I estimate the age of each star by fitting a weighted probability density function to the age distribution of the models that best match the seismic constraints.}
   {To evaluate the performance of the fitting method, it was applied to a synthetic population of 20 $\delta$ Sct stars with varying metallicities and ages, generated by randomly selecting models. The analysis indicates that $\delta$ Sct stars older than 1 Gyr, but still prior to reaching the terminal-age main sequence, can in principle be reliably age-dated. Nevertheless, when the method is applied to the observational sample, only three out of the 35 stars considered marginally exceed an estimated age of 1 Gyr.}
   {From these results, I can say that open clusters older than approximately 1 Gyr cannot be reliably dated using asteroseismology of $\delta$ Sct stars with 1D models, at least not without a more complete treatment of convection and a non-linear treatment of rotation.}


   \maketitle

\section{Introduction}\label{sec:introduction}

Within the Hertzsprung–Russell (HR) diagram, the classical instability strip (IS) encompasses several types of pulsating stars. The $\delta$ Sct  are Population I stars of spectral types A2 to F2, located near the main sequence (MS), with effective temperatures ranging from approximately 6400 K to 8600 K \citep{Breger2000,RodriguezBreger2001,Aerts2010,Uytterhoeven2011}. They occupy an intriguing region of the HR between low-mass stars with radiative cores and thin convective envelopes (M$\lesssim$2M$_{\odot}$), and more massive stars with convective cores and radiative envelopes (M$\gtrsim$2M$_{\odot}$).

The study of $\delta$ Sct stars provides valuable constraints on models of stellar structure and evolution, as physical properties such as the internal rotation and the size of the convective core play a key role in shaping the evolutionary paths of intermediate- and high-mass stars.
\citep{Maeder2009,Meynet2013}. They can exhibit dozens to hundreds of radial and non-radial pulsation modes, with periods ranging from approximately 15 minutes to 8 hours \citep{Uytterhoeven2011,Holdsworth2014}. These oscillations are excited by the $\kappa$-mechanism operating in the He II partial ionization zone \citep{Cox1963}. Typically, these stars are fast rotators \citep{Royer2007}. As they evolve, their pulsation modes shift toward lower frequencies, which in the HR diagram corresponds to a movement toward the red, or cooler, edge of the IS. In this region, $\delta$ Sct overlap with $\gamma$ Dor   variables, which pulsate at lower frequencies excited by a convective blocking mechanism \citep{Guzik2000,Dupret2005}.

Detecting modes in $\gamma$ Dor  stars is challenging due to their closely spaced frequencies, requiring long-duration, high-resolution observations, on the order of 90 days, to resolve them \citep{Kurtz2022}. The advent of space-based photometry from the Kepler \citep{Koch2010} and TESS \citep{Ricker} missions has led to substantial progress in the asteroseismology of $\gamma$ Dor  stars \citep{VanReeth2015}.

Determining the red edge of the IS remains a significant challenge, both theoretically and observationally. From a theoretical standpoint, models must incorporate time-dependent convection (TDC) to account for the coupling between oscillations and convection in the stellar envelope. Within the framework of Mixing Length Theory (MLT) \citep{Unno1967}, \citet{Xiong2001} derived a theoretical red edge for radial modes using the  non-local TDC formulation of \citet{Xiong1998}. Similarly, \citet{Griga2004} and \citet{Dupret2005} computed theoretical blue and red edges for both radial and non-radial modes, employing the TDC formalism developed by \citet{Gabriel1996}.

The location of the IS is highly sensitive to the choice of the free parameter $\alpha$ in the TDC treatment. According to \citet{Dupret2005}, values of $\alpha$ between 1.8 and 2.0 yield better agreement with observational data. Ideally, a grid of models allowing for variation in this parameter would be constructed, especially near the terminal-age main sequence (TAMS). \citet{Houdek2015} provides a comprehensive summary of recent advances in this area.

From an observational standpoint, since the release of data from the Kepler space mission, evidence has emerged for the existence of hybrid stars exhibiting pulsations characteristic of both $\delta$ Sct  and $\gamma$ Dor  stars \citep{Uytterhoeven2011}. This observational confirmation supported the theoretical predictions of \citet{Dupret2005}. These hybrid stars display high-frequency acoustic modes (p-modes), typical of $\delta$ Sct stars, alongside low-frequency buoyancy modes (g-modes), typical of $\gamma$ Dor  stars. The interaction between these mode types occurs near the boundary separating the convective core from the radiative envelope, giving rise to the phenomenon of avoided crossings, where p- and g-modes of similar frequencies couple and exchange characteristics. It should be noted that clear evidences of avoided crossings have not been detected in the observed spectra of $\delta$ Sct in the same way as in solar-like stars.

To facilitate the classification of $\delta$ Sct  and $\gamma$ Dor  stars, \citet{Griga2010} proposed a frequency cut-off at 5 d$^{-1}$. However, this limit is based more on statistical considerations than on physical theory. In \cite{Gautam2025} they show that $\delta$ Sct models near the zero-age main sequence (ZAMS) have g-modes around 20 d$^{-1}$. The classification becomes even more complex with the discovery of a significant fraction of stars located within the IS that do not exhibit any pulsations. In \citet{Murphy2019}, the edges of the IS were redefined based on the pulsator fraction in a sample of 15 000 A- and F-type stars.

In summary, the red edge of the IS is a theoretically and observationally challenging boundary to define. Around this region, two types of pulsating stars coexist, often displaying oscillation modes with similar frequencies but differing physical origins, complicating both their identification and seismic modelling. 

One of the key seismic indicators is the large frequency separation ($\Delta\nu$), defined as the difference between acoustic modes of the same spherical degree and consecutive radial orders. This parameter is directly related to the mean density and surface gravity of the star, an approach previously applied to solar-like stars.
Identifying these regular spacings between consecutive radial modes is challenging due to the presence of avoided crossings and the typically rapid rotation of $\delta$ Sct stars.
Nevertheless, studies by \citet{GH2013, GH2015, GH2017, Paparo2016,Michel2017, Bedding2020} found evidence of periodicities in the frequency spectra of $\delta$ Sct observed with CoRoT, Kepler and TESS. 

In \citet{PamosOrtega2022, PamosOrtega2023} (hereafter Paper~I and Paper~II respectively), we used the large separation observed in a sample of $\delta$ Sct to date three young open clusters with different metallicities and ages: $\alpha$ Persei, Trumpler 10, and Praesepe.

We also employed the $\nu_{\mathrm{max}}$-T$_{\mathrm{eff}}$ scaling relation proposed by \citet{Forteza2020}. However, some authors, such as \citet{Bedding2023}, have questioned the applicability of this relation to $\delta$ Sct stars. The main concern lies in the complexity of their power spectra, which do not conform well to a Gaussian profile as observed in solar-type stars with stochastically excited oscillations and a well-defined maximum. In addition, \cite{Murphy2024} analysed the $\nu_{\mathrm{max}}$–T$_{\mathrm{eff}}$ relation in the young stellar association Cep Her, and they claim that this relation is unreliable, even for stars younger than approximately 100 \myr. Nevertheless, the independent scaling relations proposed by \citet{Forteza2020} and \citet{Hasanzadeh2021} were derived from sufficiently large and homogeneous samples of pure $\delta$ Sct stars. Although the dispersion in the relation is significant, it can still serve as a useful constraint for stellar models via the effective temperature, particularly in cases where the oscillation envelope is narrow and contains tightly clustered acoustic modes. This behaviour is more typical of young $\delta$ Sct located near the ZAMS or in the early stages of the MS. In contrast, for more evolved $\delta$ Sct along the MS, the $\nu_{max}-T_{eff}$ relation becomes less reliable due to an increased dispersion. For this reason, in the present work, I choose to use available spectroscopic temperatures to constrain the models.   

Many intermediate-mass stars exhibit moderate to high rotation rates, which further complicates the identification of observed oscillation modes. Under the assumption of uniform rotation and using first-order perturbation theory, the rotational splitting of a mode frequency $\omega_{n,\ell}$ in azimuthal order $m$ is described by the equation $\omega_{n,\ell,m}=\omega_{n,\ell}+m(1 - C_{n,\ell})\Omega$, where $\Omega$ is the rotation frequency and $C_{n,\ell}$ is the Ledoux constant, accounting for the influence of the Coriolis force. Rotation thus lifts the degeneracy of the mode $\omega_{n,\ell}$, producing a multiplet of $(2\ell + 1)$ distinct frequencies corresponding to different values of $m$, for a fixed value of $\ell$. In the case of high-order p-modes, the Ledoux constant is approximately zero \citep{Aerts2010}. Under these conditions, the resulting multiplets, comprising components with
$m = -\ell, -\ell + 1, .... 0. .... \ell-1, \ell$, constitute a useful seismic observable for determining the stellar envelope's rotation frequency independently of the equilibrium model. In Paper~I and Paper~II, these frequency multiplets were also used as a seismic diagnostic to estimate the rotation rate and constrain the stellar models, in cases where such structures could be reliably detected. 

Oscillations in rapidly rotating stars cannot be accurately described using spherical harmonics. More sophisticated perturbative approaches have been developed, incorporating the Coriolis force up to second- and third-order terms in frequency calculations based on 1D equilibrium models \citep[see][for a review of the effects of rapid rotation on mode geometry]{Mirouh2022}. In Paper~I and Paper~II, we employed a perturbative method up to third order, accounting for stellar deformation and near-degeneracy effects, as developed by \citet{Suarez2002} and implemented in the \filou\ code \citep{Suarez2008}, to compute oscillation modes across our model grids. However, since these phenomena are not relevant to the present analysis, and considering the higher computational cost associated with \filou, the \gyre\ code \citep{Townsend2013} has been selected for this work as a more efficient alternative. As the present oscillation models consider only radial acoustic modes (Sect.~\ref{sec:grid}), rotational multiplets were not employed as constraints in the model fitting process.

Building upon the research presented in Paper~I and Paper~II, this work further refines the conditions under which $\delta$ Sct stars, and the open clusters that host them, can be reliably dated. I analyse the frequency content of a sample of $\delta$ Sct observed in various sectors by the Kepler and TESS missions, specifically focusing on those located near the red edge of the IS. There we can find the oldest $\delta$ Sct stars can still be on the MS, but prior to reaching the TAMS, where frequency analyses become too complex to allow reliable age determination. The primary objective is to estimate their ages and to determine an upper limit for the age of an open cluster that can be dated using $\delta$ Sct asteroseismology. 

The structure of this paper is as follows: Sect.~\ref{sec:samples} describes the selection criteria for the stellar samples. Sect.~\ref{sec:seismic} outlines the frequency analysis methodology. Sect.~\ref{sec:grid} details the construction of the model grid used in this work. Sect.~\ref{sec:sim_stars} shows the goodness of the method with a group of twenty simulated $\delta$ Sct stars. Sect.~\ref{sec:results} presents the results obtained from the statistical treatment of the models for the stellar samples. Finally, Sect.~\ref{sec:conclusions} summarizes the main conclusions of the study. 
   

\section{The samples}\label{sec:samples}

I began by identifying candidate $\delta$ Sct located as close as possible to the red edge of the IS. For this purpose, I consulted the catalogues of \citet{Murphy2019} and \citet{Stassun2019} (hereafter M2019 and S2019 respectively), which include extensive samples of stars observed by the Kepler and TESS missions, respectively. I selected stars with effective temperatures between 6500 K and 7500 K, and metallicities in the range [Fe/H] = –0.3 to +0.3.

An additional selection criterion was the availability of short-cadence light curves ($\simeq2$ minutes) in PDCSAP FLUX, from specific Kepler and TESS sectors, made accessible through the TESS Asteroseismic Science Consortium\footnote{\url{https://tasoc.dk}} (TASC). Applying these criteria resulted in a sample of 64 candidates from Kepler and 53 from TESS.

\begin{figure}   
    \centering
    \includegraphics[width=9.5cm,height=9cm,keepaspectratio]{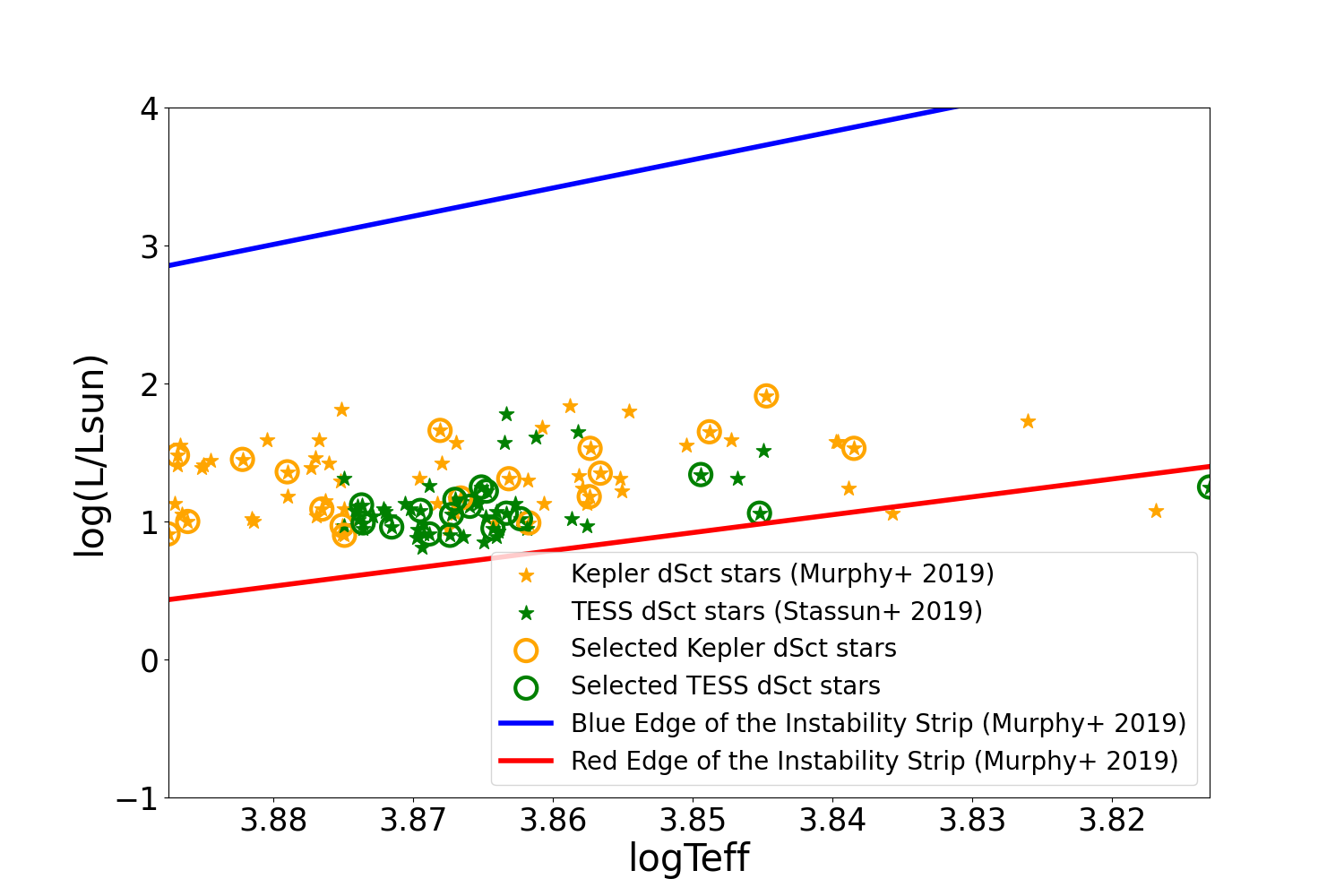}
    \caption{HR diagram showing 64 $\delta$ Sct candidates observed in various Kepler sectors (orange) and 53 $\delta$ Sct candidates observed in TESS sectors (green), all located near the red edge of the instability strip. Targets from both samples selected for seismic analysis are highlighted with circles.}
    \label{fig:HR_Kepler_TESS}
\end{figure}

The Figure~\ref{fig:HR_Kepler_TESS} shows the location of these stars in the HR diagram. For consistency with the underlying data, I adopted the IS boundaries as defined by \citet{Murphy2019}.

Once the oscillation frequencies were extracted using \multim\footnote{\url{https://github.com/davidpamos/MultiModes}}, the open-source code developed in Paper~I and publicly available, I visually inspected the periodograms of all the candidate stars. This step was crucial to discard those stars exhibiting oscillation spectra with clear signs of avoided crossings or indications of high rotational velocity, both of which tend to blend oscillation modes and hinder mode identification.

I retained only those stars whose frequency spectra displayed well-separated p-modes, clearly distinguishable from low-frequency signals typically associated with convection or rotation. These selected targets, 18 from the Kepler sample and 17 from TESS, are marked with circles in Fig.~\ref{fig:HR_Kepler_TESS}. The Tables~\ref{tb:Kepler_sample} and \ref{tb:TESS_sample} in Appendix~\ref{AppendixA} list the main parameters of each selected star in both samples.

Among the listed parameters, I highlight the projected rotational velocity, vsini, compiled from various sources. Candidates with 
\vsini>120 \kms\ and periodograms dominated by a complex mix of low and high frequencies were excluded from the final analysis, as their seismic modelling would be unreliable with the methods used here.

\begin{figure}   
    \centering
    \includegraphics[width=9cm,height=8.5cm,keepaspectratio]{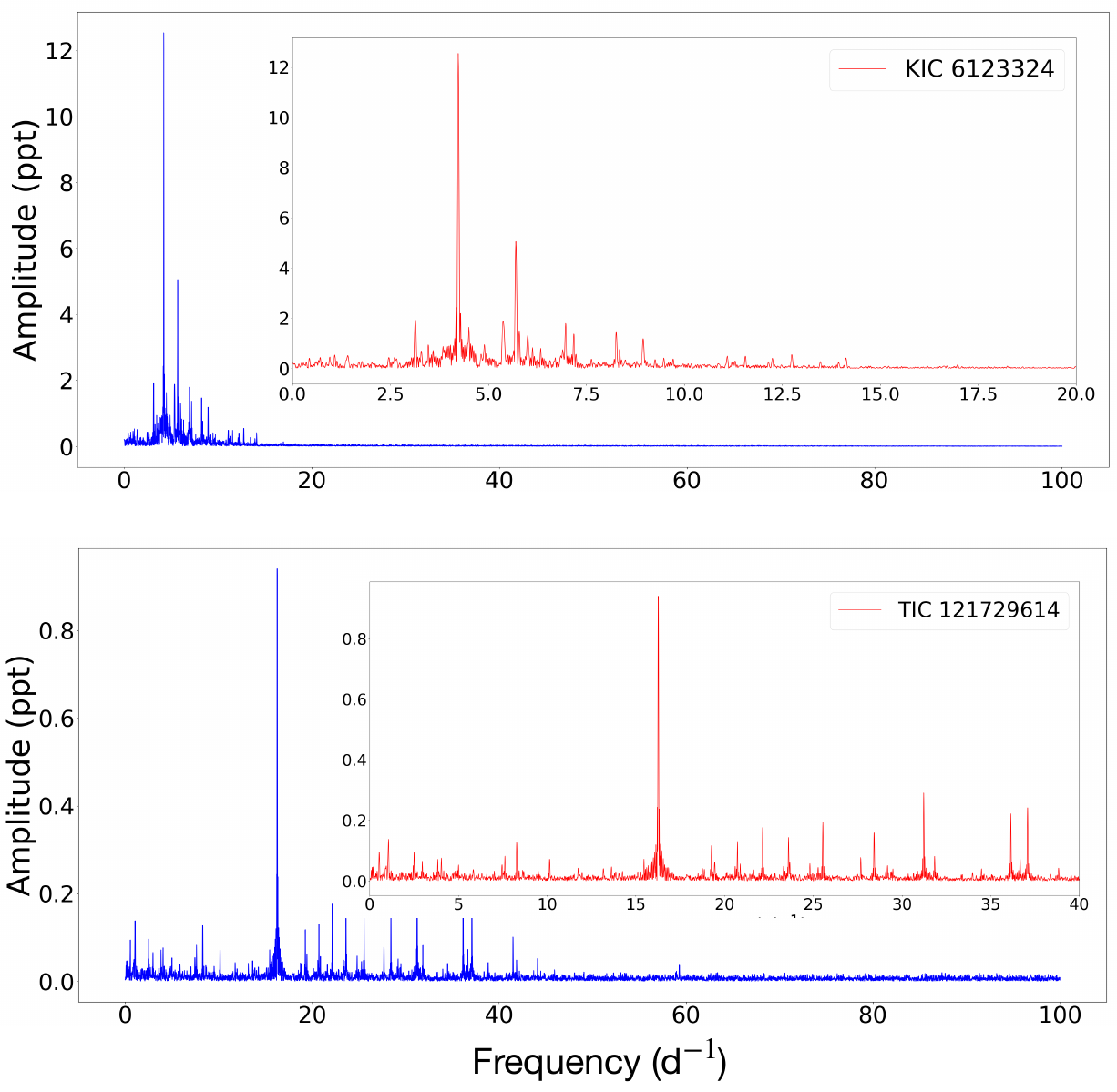}
    \caption{Top: Periodogram of KIC~6123324. Bottom: Periodogram of TIC~121729614.}
    \label{fig:Discards_Kepler_TESS}
\end{figure}

The Figure~\ref{fig:Discards_Kepler_TESS} illustrates two discarded cases. The top panel shows the periodogram for KIC~6123324, where most of the detected frequencies lie below 10 d$^{-1}$ , suggesting that rotational or convective processes dominate its variability. The bottom panel shows the periodogram of TIC~121729614, which presents a dense forest of frequencies, both low and high, along with evidence of avoided crossings. Its high projected rotational velocity, 260 \kms\ \citep{Niemczura2015}), supports this interpretation.  

Although the stars in each group may exhibit a wide range of metallicities, both samples can provide an empirical upper limit on the age at which $\delta$ Sct stars, and by extension, the open clusters that host them, can be reliably dated through asteroseismology. 

\section{The seismic analysis}\label{sec:seismic}

\begin{figure*}   
    \centering
    \sidecaption
    \includegraphics[width=12cm,height=10cm,keepaspectratio]{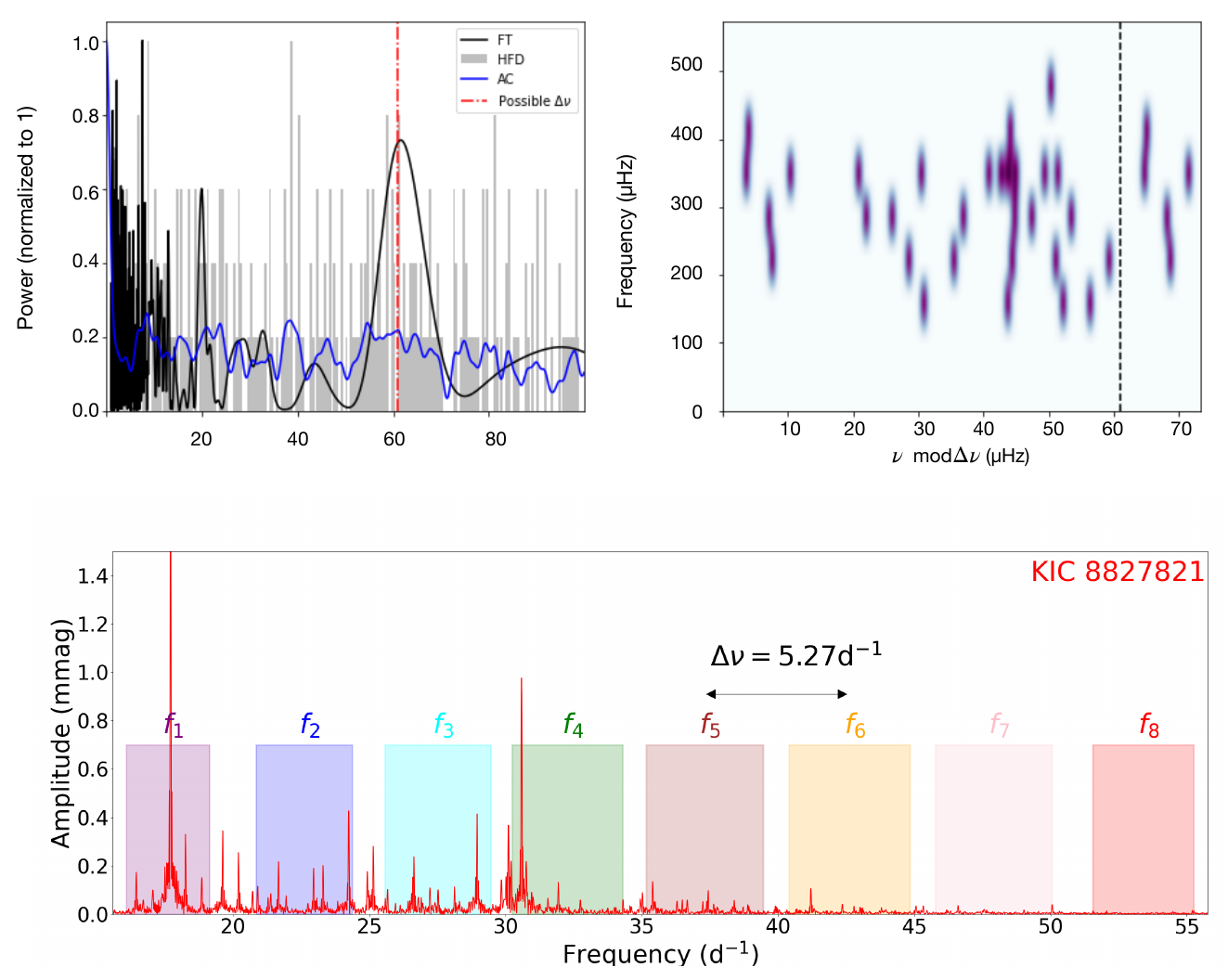}
    \caption{Top: Estimated large separation for KIC~8827821, $\Delta\nu$ = 61 $\mu$Hz = 5.27 d$^{-1}$. Top Left: Fourier Transform (FT), Autocorrelation (AC) and the Histogram of Frequency Differences (HFD). Top Right: Echélle Diagram (ED). Bottom: Frequency ranges corresponding to radial modes with radial orders $n=1$ to $n=8$, based on the estimated large separation.}
    \label{fig:dnu_KIC8827821}
\end{figure*}

After extracting the oscillation frequencies for each star in both samples using the \multim\ code, I estimated the large frequency separation ($\Delta\nu$) following the same techniques described in Paper~I and Paper~II (and references therein). As an illustrative example, the top panel of Fig.~\ref{fig:dnu_KIC8827821} shows the estimated $\Delta\nu$ for KIC~8827281. Each figure is divided into two panels: the left panel presents the $\Delta\nu$ estimation obtained using three different methods: the Fourier transform (FT), the autocorrelation function (AC), and the frequency separation histogram (HFD), while the right panel displays the corresponding échelle diagram (ED). The results for the full set of selected targets are shown in a Zenodo public repository (\href{url}{https://doi.org/10.5281/zenodo.17067341}).

In addition to the large separation, I used the available spectroscopic temperature, taken from the M2019 catalogue for the Kepler sample and from the S2019 catalogue for the TESS sample, to constrain the stellar models computed in this work. These constraints allowed me to compare the observed frequencies with those predicted for radial modes with orders $n=1$ to $n=8$. The objective was to define a frequency range for each radial mode within the observed spectrum. This approach provides greater confidence in the $\Delta\nu$ estimates and allows for tighter constraints on stellar ages in the next step of the analysis.

The bottom panel of Fig.~\ref{fig:dnu_KIC8827821} illustrates this method: coloured rectangles indicate the frequency ranges in which each of the first eight radial modes is expected to lie, based on the estimated $\Delta\nu$ for KIC~8827821. 

The corresponding radial mode frequency ranges for all the selected stars in the Kepler and TESS samples are available in \href{url}{https://doi.org/10.5281/zenodo.17067341}. The specific numerical values used in this analysis are also provided in Table~\ref{tb:seismic_parameters_Kepler} and Table~\ref{tb:seismic_parameters_TESS} in Appendix~\ref{AppendixB}.

\section{The grid}\label{sec:grid}

To characterise $\delta$ Sct stars in the MS, I computed a grid of stellar models using \mesa, version r24081 \citep{Paxton2019}, evolving each model up to the TAMS. As configured by default, the code was allowed to determine the optimal time step between consecutive models during the evolution. The models were calculated based on the set of independent parameters listed in Table~\ref{tb:grid}. The nuclear reaction network used was pp$\_$and$\_$cno$\_$extras.net, following the prescriptions of \citet{Murphy2023}.

At the ZAMS, the initial chemical composition satisfies X + Y + Z = 1. For a fixed value of hydrogen mass fraction X, increasing metallicity Z results in a corresponding decrease in helium fraction Y. The models incorporate differential rotation, with rotation initiated at the ZAMS. Initial angular velocity to critical velocity ratios were chosen in the range $0.1\Omega_{c}$ to $0.5\Omega_{c}$, avoiding higher values that exceed the applicability limits of the perturbative approach and adhering to the assumptions underlying the \gyre\ pulsation code.

Convective core overshooting was not included. In Paper~II, we tested exponential overshooting values of f$_{0}$ = 0.002 and f = 0.022, and found no significant impact on the age uncertainties of Trumpler~10 and Praesepe. In the present study, the vast majority of stars in the sample, if not all, have not yet reached the TAMS, as inferred from their positions in the HR diagram, where the effects of the overshooting become significant. Moreover, the primary objective of this work is not to provide a precise characterization of each individual star. 

The atmospheric boundary condition was set using \mesa’s default configuration, adopting a fixed Eddington 
T-$\tau$ relation \citep{Eddington1926}. The choice of an atmospheric model has a negligible impact on the oscillation frequencies considered here, which are excited at sufficiently deep layers within the radiative envelope.

I computed the oscillation modes using \gyre, version 7.2.1. As explained in Sect.~\ref{sec:samples}, the focus of this study is on stars that predominantly exhibit p-modes rather than g-modes. Accordingly, I limited the computation to the first eight radial p-modes. In this regime, the Coriolis force has a negligible effect on the eigenfunctions of the modes, and the influence of centrifugal distortion is even smaller. Under these conditions, the use of 1D oscillation models, enabled by the close coupling between the equilibrium models computed with \mesa\ and the oscillation models computed with \gyre, is fully justified. 

The oscillation computations employed the second-order Gauss–Legendre Magnus integration scheme (MAGNUS$\_$GL2). I performed frequency scans in the range 5 to 100 d$^{-1}$ , using a grid of 100 frequency points. To reproduce the expected distribution of p-modes, I selected a linear frequency grid (grid$\_$type = 'LINEAR'), appropriate for modes that are approximately equally spaced in frequency at high radial orders. It is well established that radial modes are not strictly equally spaced, with the first four modes typically lying closer together than the subsequent four. Accordingly, the large frequency separation was determined as the average of the separations between consecutive radial modes in the range $n=5$ to $n=8$. For refining the mesh of integration points around each trial frequency, I adopted the recommended \gyre\ parameters: w$_{osc}$ = 10, w$_{exp}$ = 2 and w$_{ctr}$ = 10.

\begin{table}
	\centering
	\caption{Parameters of the stellar model grids computed with the \mesa\ code.}
    \renewcommand{\arraystretch}{1.5}
	\addtolength{\tabcolsep}{8 pt}
	\resizebox{7cm}{!}{
    \begin{tabular}{ccc}
			\hline
			Parameter & Range & Step\\
			\hline
			M ({\rm M}$_{\odot}$) & [1.56; 2.44] & 0.02 ${\rm M}_\odot$\\
			Z & [0.010; 0.030] & 0.002\\
			$\Omega / \Omega_{c}$ & [0.1; 0.5] & 0.2\\
			$\alpha$ & 1.9 & Fixed\\
			
    \end{tabular}
    }
    \tablefoot{From top to bottom: stellar mass, metallicity, initial angular velocity to critical velocity ratio, and mixing length parameter.}
	\label{tb:grid}
\end{table}

\section{Testing the method with a group of 20 simulated $\delta$ Sct stars}\label{sec:sim_stars}
The statistical treatment applied here follows the same methodology as in Paper~II. As a reminder, a major source of bias in the computed models arises from the oversampling of very young models produced by \mesa. This effect is mitigated by applying a weighting factor proportional to the time step between consecutive models along the evolutionary track. The impact is most relevant near the PMS and the ZAMS, while it becomes negligible for more evolved models on the MS.

Using the updated Weighted Probability Density Function (WPDF), I have computed histograms of the constrained models and fitted a Gaussian distribution to them, evaluating the goodness of fit using the Pearson chi-squared statistic, $\chi^2$. 

Prior to applying the method to the observational samples, its reliability was assessed using simulated stars. Twenty models were randomly selected from the previously computed grid (Table~\ref{tb:Parameters_simulated_stars}), covering effective temperatures between 6500 K and 7500 K, and spanning a range of metallicities and ages.

\begin{table*}
    
    \centering
	\caption{Parameters of 20 simulated $\delta$ Sct stars.}
    \renewcommand{\arraystretch}{1.5}
	\addtolength{\tabcolsep}{2.5 pt}
	\resizebox{\textwidth}{!}{
    \begin{tabular}{ccccccccccc}
			\hline
			ID & Z & $\mathrm{log\ (L/L_{\odot})}$ & $\mathrm{M\ (M_{\odot})}$ & $\mathrm{R\ (R_{\odot})}$ & $\bar \rho\ (\bar \rho_{\odot})$ & \logg & $\mathrm{T_{eff}}$ (K) & v (\kms) & $\Delta\nu$ (d$^{-1}$) & Age (\myr) \\
			\hline
	       	1 & 0.012 & 1.3 & 1.88 & 2.72 & 0.09 & 3.84 & 7401 & 94 & 39 & 958 \\
			2 & 0.010 & 1.13 & 1.70 & 2.33 & 0.13 & 3.93 & 7239 & 168 & 45 & 1125 \\
            3 & 0.014 & 0.88 & 1.56 & 1.65 & 0.35 & 4.20 & 7454 & 125 & 73 & 775 \\
            4 & 0.014 & 1.21 & 1.82 & 2.45 & 0.12 & 3.92 & 7401 & 34 & 45 & 989 \\
            5 & 0.016 & 1.39 & 2.02 & 2.96 & 0.08 & 3.80 & 7480 & 30 & 36  & 839\\
            6 & 0.016 & 0.89 & 1.60 & 1.73 & 0.31 & 4.16 & 7314 & 204 & 67 & 630 \\
            7 & 0.018 & 0.87 & 1.58 & 1.71 & 0.32 & 4.17 & 7293 & 42 & 71 & 829 \\
            8 & 0.016 & 0.90 & 1.62 & 1.71 & 0.32 & 4.18 & 7436 & 207 & 524 \\
            9 & 0.020 & 1.18 & 1.86 & 2.62 & 0.10 & 3.87 & 7046 & 32 & 41 & 1002 \\
            10 & 0.020 & 1.37 & 2.06 & 2.93 & 0.08 & 3.82 & 7408 & 94 & 36 & 776 \\
            11 & 0.022 & 1.34 & 2.06 & 2.79 & 0.09 & 3.86 & 7483 & 164 & 38 & 711 \\
            12 & 0.020 & 1.16 & 1.84 & 2.55 & 0.11 & 3.89 & 7034 & 98 & 42 & 999 \\
            13 & 0.022 & 1.31 & 1.88 & 2.80 & 0.09 & 3.82 & 7337 & 92 & 37 & 1135 \\
            14 & 0.022 & 0.98 & 1.74 & 1.84 & 0.28 & 4.15 & 7483 & 206 & 63 & 424 \\
            15 & 0.026 & 1.45 & 2.16 & 3.36 & 0.06 & 3.72 & 7255 & 86 & 30 & 776 \\
            16 & 0.026 & 0.96 & 1.68 & 2.22 & 0.15 & 3.97 & 6722 & 36 & 51 & 1183 \\
            17 & 0.026 & 1.01 & 1.76 & 1.97 & 0.23 & 4.09 & 7349 & 122 & 60 & 664 \\
            18 & 0.028 & 1.39 & 1.98 & 3.05 & 0.07 & 3.77 & 7356 & 30 & 34 & 1010 \\
            19 & 0.030 & 1.41 & 2.02 & 3.33 & 0.05 & 3.70 & 7130 & 139 & 29 & 981 \\
            20 & 0.030 & 1.53 & 2.32 & 3.65 & 0.05 & 3.68 & 7301 & 140 & 27 & 624 \\

    \end{tabular}
    }
    \tablefoot{From left to right: Identifier number (ID), metallicity (Z), luminosity (log(L/L$_\odot$)), mass (M/M$_\odot$), radius (R/R$_\odot$), mean density ($\bar{\rho}/\bar{\rho}\odot$), surface gravity (\logg), effective temperature (T$_\mathrm{eff}$), rotational velocity (v$_{\mathrm{rot}}$), large frequency separation ($\Delta\nu$) and age.}
	\label{tb:Parameters_simulated_stars}
\end{table*}

\begin{figure*}   
    \centering
    \includegraphics[width=18cm,height=22cm,keepaspectratio]{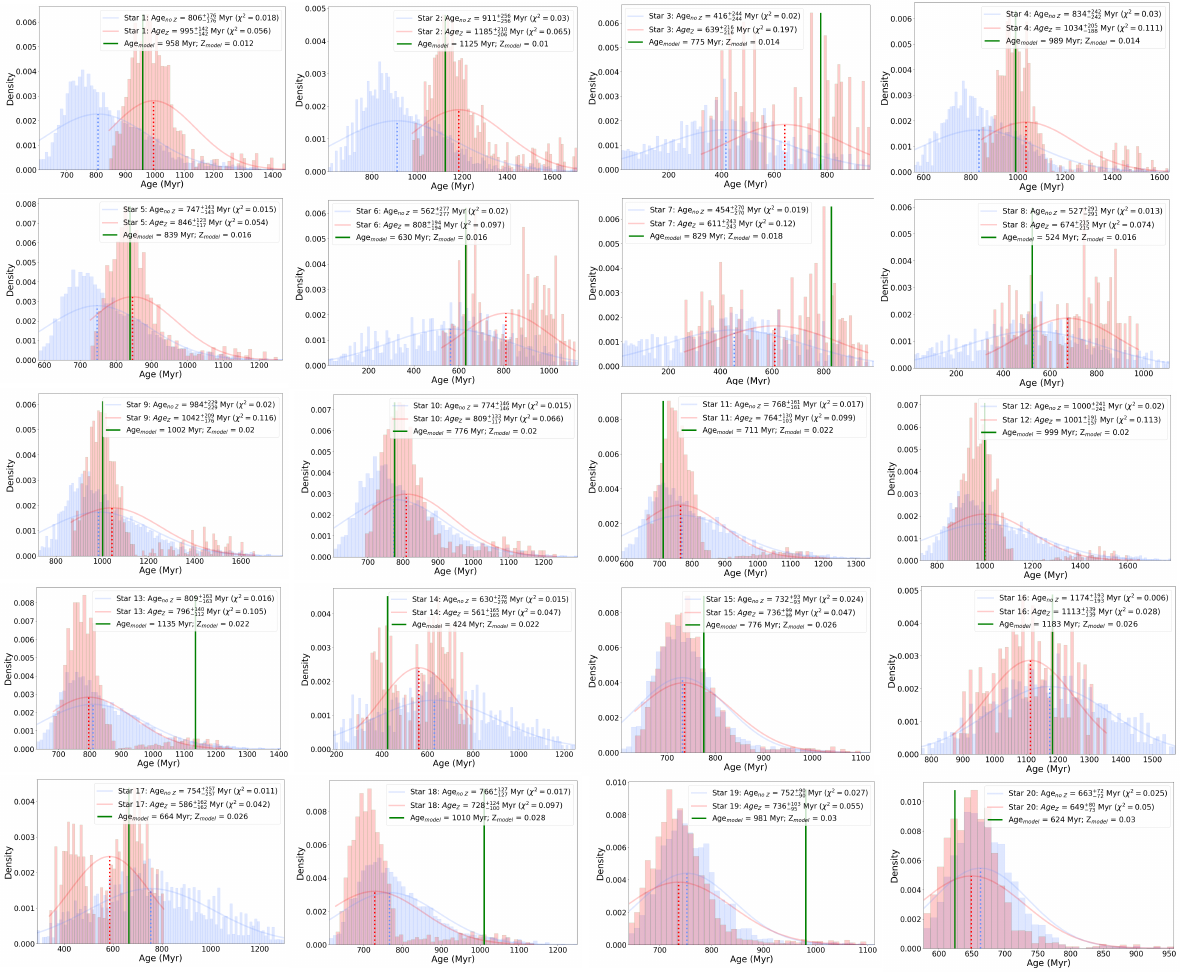}
    \caption{Histograms and WPDFs of twenty simulated stars, with different metallicities: when Z is used (red) and not used (blue) as an observational constraint. The green solid line represents the real age of the model.}
    \label{fig:simulated_stars}
\end{figure*}

For each simulated star, the effective temperature and the large frequency separation were adopted as observational constraints. In all cases, the effective temperature was assigned an uncertainty of 150 K, a value representative of many stars in the Kepler and TESS samples. For the large frequency separation, an uncertainty of 2 $\mu$Hz was assumed, corresponding to the average uncertainty measured in real stars. Treating the objects as field stars, metallicity was not included as a constraint, since this parameter is typically subject to large uncertainties. Conversely, if the stars are assumed to belong to a cluster of known metallicity, this parameter can be incorporated to further restrict the models. For each case, histograms and WPDFs were computed (Figure~\ref{fig:simulated_stars}). The resulting fits, those constrained by metallicity (red) and those unconstrained (blue), were compared with the true age of the simulated star (green line). This procedure provides an estimate of the intrinsic uncertainty associated with the fitting method.

Several conclusions can be drawn from this study. First, as expected, the distributions are more Gaussian-like when metallicity is not used as a constraint. Second, in only eight stars (numbers 1, 2, 3, 4, 5, 8, 14, and 20), constraining the models by metallicity yields ages closer to the true values, primarily in the subsolar regime. However, the corresponding histograms do not exhibit Gaussian distributions, owing to the insufficient number of models, either because the stellar mass lies near the lower limit of the grid, or because the rotation rate approaches the upper boundary. Third, the fits obtained for solar metallicity are very similar, as expected, given that the grid was designed with Z$\simeq$0.020 as the average value. Finally, for three of the 20 stars (numbers 13, 18, and 19), no reliable age determination was possible within the adopted uncertainties. Notably, these correspond to some of the oldest stars in the sample. Near the TAMS, the grid density becomes insufficient to achieve a good fit consistent with a normal distribution. Nevertheless, stars numbered 2, 9, and particularly 16—each older than 1 Gyr—have been dated with remarkable precision. This result demonstrates that, within this grid, reliable age determinations can be extended to stars well beyond this threshold. 

This simulation supports a key conclusion of the study: metallicity should not be used as a constraint in the modelling of individual stars with the present grid. It is not a reliable observable, as its reported values vary significantly across different sources and catalogues. In addition, the number of models in the grid becomes insufficient near the extreme values, preventing the construction of normal distributions that can be reliably fitted. A more finely sampled metallicity grid would therefore be required for modelling individual stars, or in cases where the objective is to determine the age of a cluster. 

\section{Results and discussion}\label{sec:results}

\begin{figure*}  
    \centering
    \includegraphics[width=16cm,height=12cm,keepaspectratio]{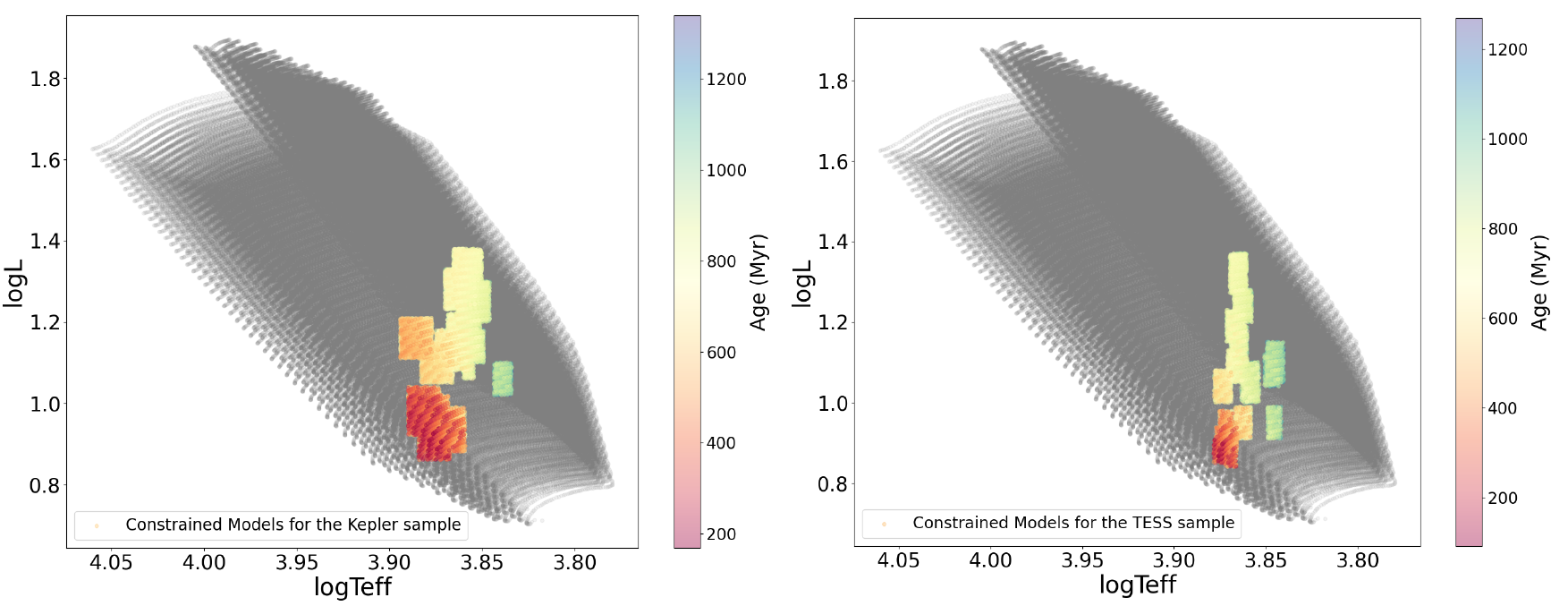}
    \caption{Left panel: Seismically constrained stellar models for the Kepler sample of $\delta$ Sct stars, using large separation, spectroscopic temperatures, and metallicities. Right panel: Same as left, but for the TESS\ sample.}
    \label{fig:HR_constrained_models}
\end{figure*}

\begin{figure}   
    \centering
    \includegraphics[width=9cm,height=7cm,keepaspectratio]{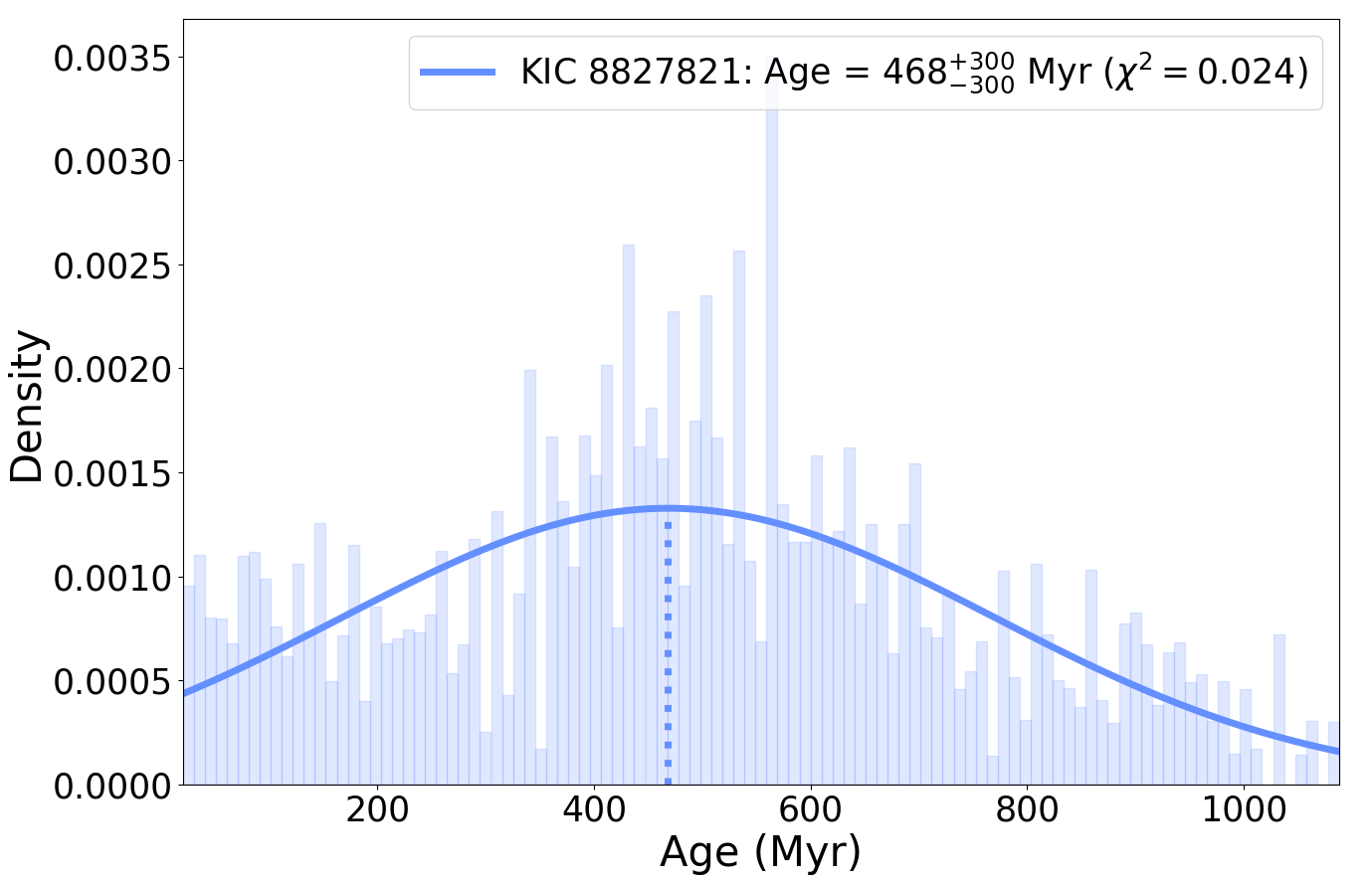}
    \caption{Age histogram and Weighted Probability Density Function (WPDF) derived from the seismically constrained models for KIC~8827821. The Gaussian fit and the corresponding $\chi^2$ value are also shown.}
    \label{fig:histogram}
\end{figure}

The Figure~\ref{fig:HR_constrained_models} shows the constrained models in $\Delta\nu$ and T$_{eff}$ for the Kepler (left panel) and the TESS (right panel) samples. And the Figure~\ref{fig:histogram} show the age histogram, the WPDF and the estimated age for KIC~8827821. The complete set of histograms for the Kepler and TESS\ samples is available in \href{url}{https://doi.org/10.5281/zenodo.17067341}. The parameters corresponding to each of the modelled stars are shown in Tables\ref{tb:Parameters_models_Kepler} and \ref{tb:Parameters_models_TESS} in the Appendix~\ref{AppendixC}.

The ages of the Kepler sample range from $404\pm236$ \myr, corresponding to the star KIC~10686752, to $1095\pm246$ \myr, corresponding to KIC~7106205. For the TESS sample, the range spans from $331\pm240$ \myr\ (TIC~69025963) to $1029\pm241$ \myr\ (TIC~398733851). The models are consistent in metallicity, with all of them exhibiting values around Z = 0.020–0.022. This uniformity near the solar value results from the computed models spanning Z = 0.010–0.030. It is likely that several stars have not been accurately modelled, given the expected diversity in metallicity. Nevertheless, this does not significantly affect the estimated average age for each sample, as I have shown in Sect.~\ref{sec:sim_stars}. 

\begin{figure*}   
    \centering
    \includegraphics[width=16cm,height=10cm,keepaspectratio]{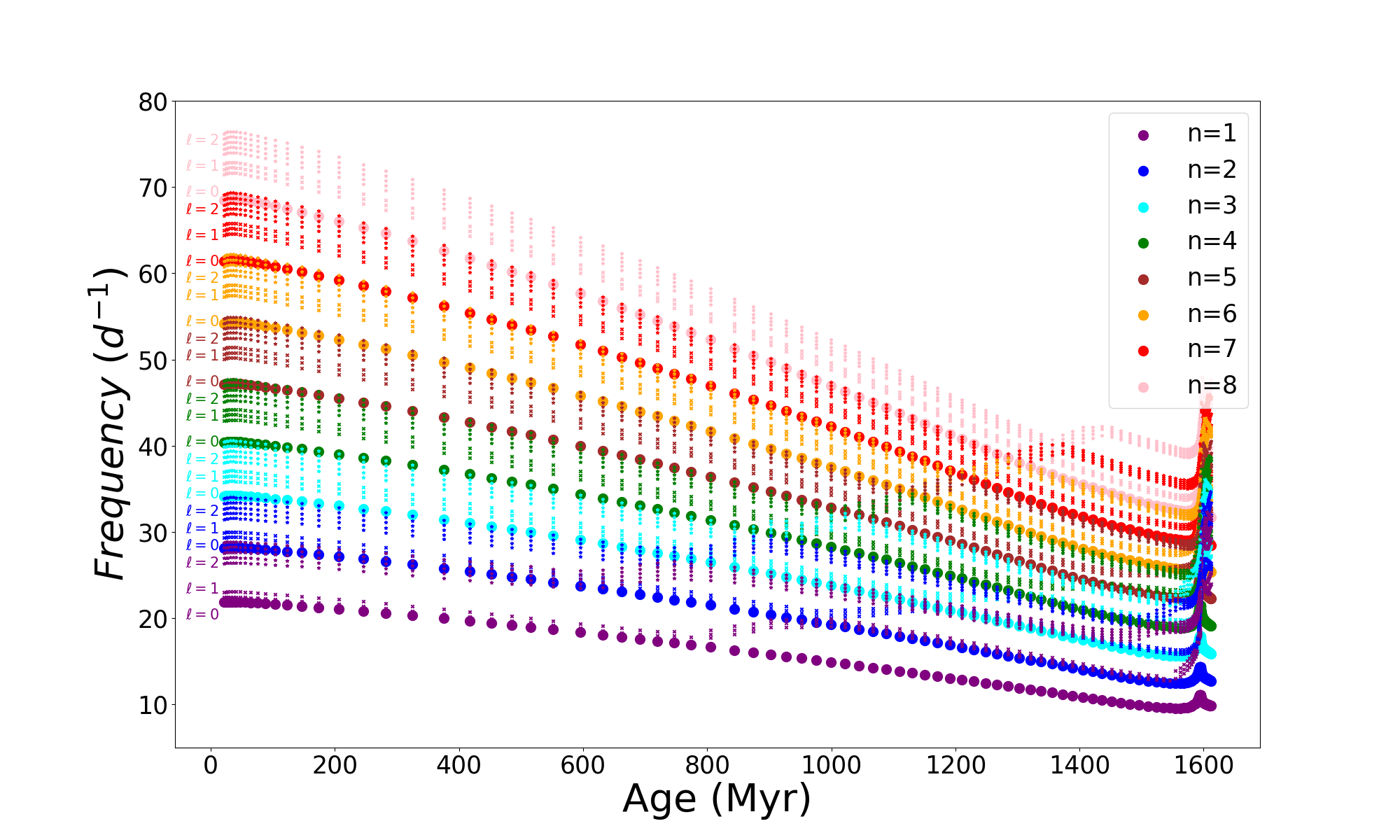}
    \caption{Evolution of the oscillation mode frequencies along the MS, for radial orders between $n = 1$ and $n = 8$, and degrees $\ell = 0,\ 1,\ 2$, corresponding to an evolutionary track with M = 1.64 M$_{\odot}$, metallicity Z = 0.018, and a rotation rate of $\Omega/\Omega_{\mathrm{crit}} = 0.10$.}
    \label{fig:Evolution_modes}
\end{figure*}

With these results, I can conclude that the oldest stars that can be reliably dated using this technique, at least not with 1D models, without a more complete treatment of convection and a non-linear treatment of rotation, barely reach 1000 \myr. To further support this claim, in the Fig.~\ref{fig:Evolution_modes}, I show the MS evolution of radial modes ($\ell = 0$) and non-radial modes ($\ell = 1,\ 2$) for an evolutionary track with M = 1.64 M$_{\odot}$, solar-like metallicity (Z = 0.018), and $\Omega/\Omega_{\mathrm{crit}} = 0.10$, representative values of the Kepler sample. As age increases beyond 600 \myr, non-radial modes of different orders begin to interact and overlap. Around 1000 \myr, this mixing extends to fifth-order modes, significantly affecting the determination of the large separation. In addition to the effects of rotation and the avoided crossings phenomenon, where p-modes mix with g-modes, the overlapping of p-modes themselves further complicates the estimation of the large separation. Although the sample is not large enough to draw a definitive statistical limit, I think that these factors are sufficient to prevent old $\delta$ Sct from being dated, since the large separation becomes indeterminable, and mode identification in the frequency spectrum is no longer feasible. 

\section{Conclusions}\label{sec:conclusions}

One-dimensional stellar models that neglect rapid rotation and lack a proper treatment of convection, such as those used in the present research, are insufficient for reliably determining the ages of $\delta$ Sct stars older than 1 Gyr. A statistical analysis performed on a synthetic stellar sample suggests that such age determination could, in principle, be achieved, if a sufficiently large number of models is available to yield a normally distributed histogram. However, when applied to the observed sample, only three stars, KIC~7106205, TIC~394015973 and TIC~398733851 exceed this age threshold. This limitation is likely attributable to the inability to obtain reliable estimates of the large frequency separation for stars excluded during the initial selection process.

\section*{Data Availability}
Plots of the large frequency separation estimates, radial mode ranges, and age histograms for each star are publicly available at Zenodo (\href{url}{https://doi.org/10.5281/zenodo.17067341}).

\begin{acknowledgements}
I appreciate the comments and questions from the anonymous referee, because they have contributed to improving the paper. 
I also appreciate the work of Dr. Juan Carlos Suárez and Dr. Antonio García Hernández, from the University of Granada, who were my thesis project directors, which I completed in December 2024. 

\end{acknowledgements}

\bibliographystyle{aa}
\bibliography{bibl}    
\clearpage

\onecolumn
\begin{appendices}

\setcounter{table}{0} 
\renewcommand{\thetable}{A.\arabic{table}}

\section{Parameters of the Kepler and TESS samples}
\label{AppendixA}

\begin{table*}[h!]
\centering
    
	\caption{Parameters of the  Kepler sample of $\delta$ Sct stars.}
    \label{tb:Kepler_sample}
    \renewcommand{\arraystretch}{1.5}
	\addtolength{\tabcolsep}{2.5 pt}
	\resizebox{\textwidth}{!}{
    \begin{tabular}{ccccccccccc}
			\hline
			KIC & G$_{mag}^{(1)}$ & $\mathrm{log\ (L/L_{\odot})}^{(1)}$ & $\mathrm{M\ (M_{\odot})^{(1)}}$ & $\mathrm{R\ (R_{\odot})^{(1)}}$ & $\bar \rho\ (\bar \rho_{\odot})$ & \logg$^{(1)}$ & $\mathrm{T_{eff}}$ (K)$^{(1)}$ & \vsini\ (\kms) & [Fe/H]$^{(1)}$  & Spectral type \\
			\hline
            3219256  & 8.279 & $0.97\pm0.021$ & $1.53\pm0.25$ & $1.81\pm0.09$ & $0.258\pm0.081$ & $4.11\pm0.08$ & $7500\pm150$ & $96^{4}$ &-0.02 &F0IV$^{(2)}$ \\
			3760002  & 10.632 & $1.13\pm0.040$ & $1.63\pm0.25$ & $2.32\pm0.19$ & $0.131\pm0.052$ & $3.92\pm0.10$ & $7255\pm253$ & - &-0.12 & F1V$^{(2)}$ \\
            3942911 & 10.837 & $1.04\pm0.022$ & $1.58\pm0.25$ & $1.95\pm0.14$ & $0.213\pm0.080$ & $4.06\pm0.09$ & $7532\pm263$ & $93^{4}$ & 0.03 & F0V$^{(2)}$ \\
            5428254 & 10.506 & $1.02\pm0.026$ & $1.58\pm0.25$ & $1.87\pm0.15$ & $0.242\pm0.096$ & $4.09\pm0.10$ & $7612\pm266$ & $96^{4}$ & 0.13 & - \\
            5474427 & 11.443 & $1.33\pm0.034$ & $1.80\pm0.25$ & $2.96\pm0.21$ & $0.069\pm0.024$ & $3.75\pm0.09$ & $7213\pm220$ & $<120^{(2)}$ & -0.07 & F0V$^{(2)}$ \\
            5774557 & 11.056 & $1.30\pm0.019$ & $1.78\pm0.25$ & $2.82\pm0.20$ & $0.079\pm0.028$ & $3.79\pm0.09$ & $7274\pm254$ & $96^{4}$ & -0.07 &F0IV$^{(2)}$ \\
            7106205 & 11.514 & $1.24\pm0.018$ & $1.72\pm0.25$ & $2.90\pm0.14$ & $0.071\pm0.020$ & $3.74\pm0.08$ & $6900\pm138$ & $<120^{2}$ & 0.30 &F1V$^{(2)}$ \\
            7217483 & 10.702 & $1.31\pm0.037$ & $1.78\pm0.25$ & $2.92\pm0.24$ & $0.071\pm0.028$ & $3.76\pm0.10$ & $7165\pm250$ & $150^{2}$ & -0.10 & F0IV$^{(2)}$ \\
            7548479 & 8.347 & $0.91\pm0.024$ & $1.49\pm0.25$ & $1.68\pm0.08$ & $0.314\pm0.098$ & $4.17\pm0.09$ & $7500\pm150$ & $10^{5}$ & 0.01 & A4V$^{(6)}$ \\
            8103917 & 11.561 & $1.17\pm0.020$ & $1.68\pm0.25$ & $2.38\pm0.17$ & $0.125\pm0.045$ & $3.91\pm0.09$ & $7355\pm257$ & - & 0.05 & Am$^{(2)}$ \\
            8446738 & 11.089 & $0.95\pm0.034$ & $1.50\pm0.25$ & $1.82\pm0.15$ & $0.249\pm0.103$ & $4.09\pm0.10$ & $7370\pm257$ & - & -0.20 & A7III$^{(2)}$ \\
            8827821 & 11.160 & $1.00\pm0.026$ & $1.56\pm0.25$ & $1.83\pm0.14$ & $0.255\pm0.099$ & $4.11\pm0.10$ & $7610\pm266$ & - & -0.01 &A8IV-V$^{(3)}$ \\
            9489590 & 10.906 & $1.13\pm0.040$ & $1.64\pm0.25$ & $2.23\pm0.20$ & $0.148\pm0.062$ & $3.95\pm0.10$ & $7383\pm258$ & - & 0.03 &A9III$^{(2)}$ \\
            9762713 & 11.206 & $1.35\pm0.017$ & $1.82\pm0.25$ & $3.05\pm0.19$ & $0.064\pm0.021$ & $3.73\pm0.08$ & $7188\pm210$ & - & 0.17 & - \\
            10549371 & 9.469 & $1.18\pm0.020$ & $1.67\pm0.25$ & $2.49\pm0.08$ & $0.108\pm0.027$ & $3.87\pm0.07$ & $7201\pm80$ & $71^{3}$ & 0.07 & F2II$^{(2)}$ \\
            10686752 & 11.311 & $0.90\pm0.044$ & $1.49\pm0.25$ & $1.66\pm0.15$ & $0.326\pm0.143$ & $4.17\pm0.11$ & $7497\pm262$ & - & -0.12 &A9V$^{(2)}$ \\
            11183539 & 10.769 & $1.00\pm0.018$ & $1.57\pm0.25$ & $1.79\pm0.13$ & $0.274\pm0.103$ & $4.13\pm0.09$ & $7693\pm269$ & $96^{4}$ & -0.07 & - \\
            12647070 & 10.684 & $1.09\pm0.023$ & $1.62\pm0.25$ & $2.07\pm0.16$ & $0.183\pm0.071$ & $4.01\pm0.09$ & $7498\pm262$ & - & -0.19 & - \\

    \end{tabular}
    }
    \tablefoot{From left to right: KIC identifier, apparent magnitude in the Gaia G Band (G$_{mag}$), luminosity (log(L/L$_\odot$)), mass(M/M$_\odot$), radius (R/R$_\odot$), density ($\bar{\rho}/\bar{\rho}\odot$), surface gravity (\logg), effective temperature (T$_\mathrm{eff}$), projected velocity (\vsini), metallicity ([Fe/H]) and spectral type.} 
    \tablebib{$^{(1)}$\citet{Murphy2019},
    $^{(2)}$\citet{Frasca2016},
    $^{(3)}$\citet{Niemczura2015},
    $^{(4)}$\citet{Jonsson2020},
    $^{(5)}$\citet{Gebran2016},
    $^{(6)}$\citet{Balona2015a}.}
	
\end{table*}

\begin{table*}[h!]
\centering
    \caption{Same as Table~\ref{tb:Kepler_sample} for the TESS\ sample of $\delta$ Sct stars.}
    \label{tb:TESS_sample}
    \renewcommand{\arraystretch}{1.5}
	\addtolength{\tabcolsep}{2.5 pt}
	\resizebox{\textwidth}{!}{
    \begin{tabular}{ccccccccccc}  
    \hline
    TIC & G$_{mag}^{(1)}$ & $\mathrm{log\ (L/L_{\odot})}^{(1)}$ & $\mathrm{M\ (M_{\odot})^{(1)}}$ & $\mathrm{R\ (R_{\odot})^{(1)}}$ & $\bar \rho\ (\bar \rho_{\odot})$ & \logg$^{(1)}$ & $\mathrm{T_{eff}}$ (K)$^{(1)}$ & \vsini\ (\kms) & [Fe/H]$^{(1)}$  & Spectral type\\
    \hline
    2096047  & 11.190 & $1.02\pm0.037$ & $1.62\pm0.27$ & $2.07\pm0.12$ & $0.18\pm0.05$ & $4.02\pm0.09$ & $7222\pm143$ & - & 0.09 & - \\
    3697787  & 10.640 & $0.91\pm0.049$ & $1.69\pm0.29$ & $1.74\pm0.11$ & $0.32\pm0.08$ & $4.18\pm0.09$ & $7393\pm107$ & - & 0.06 & - \\
    37908913 & 11.329 & $0.96\pm0.038$ & $1.70\pm0.28$ & $1.81\pm0.09$ & $0.29\pm0.07$ & $4.15\pm0.09$ &  $7439\pm110$ & - & -0.23 & - \\
    69025963 & 11.126 & $0.97\pm0.025$ & $1.73\pm0.28$ & $1.82\pm0.07$ & $0.29\pm0.06$ & $4.16\pm0.08$ & $7497\pm103$ & - & -0.08 & - \\
    86893888  & 10.877 & $0.81\pm0.021$ & $1.69\pm0.27$ & $1.54\pm0.06$ & $0.46\pm0.10$ & $4.29\pm0.08$ & $7402\pm100$ & - & 0.10 &A7$^{(5)}$ \\
    121597646 & 10.631 & $1.03\pm0.022$ & $1.66\pm0.28$ & $2.04\pm0.09$ & $0.20\pm0.04$ & $4.04\pm0.09$ & $7325\pm136$ & - & -0.09 & - \\
    122069152 & 10.520 & $0.99\pm0.021$ & $1.72\pm0.28$ & $1.87\pm0.08$ & $0.26\pm0.06$ & $4.13\pm0.08$ & $7474\pm145$ & - & -0.03 & A7III$^{(2)}$ \\
    155457396 & 11.212 & $1.16\pm0.052$ & $1.67\pm0.27$ & $2.34\pm0.16$ & $0.13\pm0.04$ & $3.92\pm0.10$ & $7362\pm127$ & - & -0.09 & - \\
    158787200 & 11.292 & $1.13\pm0.069$ & $1.65\pm0.28$ & $2.30\pm0.21$ & $0.14\pm0.04$ & $3.93\pm0.11$ & $7289\pm145$ & - & 0.20 & - \\
    231148059 & 11.200 & $0.89\pm0.040$ & $1.67\pm0.28$ & $1.73\pm0.10$ & $0.32\pm0.08$ & $4.19\pm0.09$ & $7352\pm149$ & - & 0.05 & A5/7$^{(6)}$ \\
    232064019 & 9.208 & $1.06\pm0.022$ & $1.54\pm0.26$ & $2.29\pm0.09$ & $0.13\pm0.03$ & $3.91\pm0.08$ & $7002\pm106$ & - & 0.12 &A9IV-V$^{(3)}$ \\
    255733180 & 9.594 & $1.25\pm0.027$ & $1.66\pm0.27$ & $2.60\pm0.12$ & $0.09\pm0.02$ & $3.83\pm0.09$ & $7330\pm135$ & - & -0.14 & - \\
    316826948 & 10.987 & $0.95\pm0.019$ & $1.72\pm0.28$ & $1.79\pm0.06$ & $0.30\pm0.06$ & $4.17\pm0.08$ & $7474\pm107$ & - & 0.09 & - \\
    387226824 & 10.244 & $1.03\pm0.034$ & $1.71\pm0.27$ & $1.96\pm0.09$ & $0.23\pm0.05$ & $4.08\pm0.08$ & $7443\pm100$ & - & 0.13 & - \\
    394015973 & 8.152 & $1.31\pm0.023$ & $1.55\pm0.25$ & $3.05\pm0.12$ & $0.05\pm0.01$ & $3.66\pm0.08$ & $7028\pm113$ & - & 0.17 & A9III-IV$^{3}$ \\
    397012320 & 9.833 & $1.02\pm0.024$ & $1.64\pm0.28$ & $2.02\pm0.09$ & $0.20\pm0.05$ & $4.04\pm0.09$ & $7283\pm129$ & - & -0.16 & A8V$^{4}$ \\
    398733851 & 9.845 & $1.51\pm0.033$ & $1.54\pm0.26$ & $3.86\pm0.20$ & $0.03\pm0.01$ & $3.45\pm0.09$ & $6998\pm129$ & - & 0.12 & A7III$^{4}$ \\
    \end{tabular}
    }
    \tablebib{$^{(1)}$\citet{Stassun2019},
    $^{(2)}$\citet{Frasca2016},
    $^{(3)}$\citet{Houk1975},
    $^{(4)}$\citet{Houk1999},
    $^{(5)}$\citet{Hou2015},
    $^{(6)}$\citet{Karlsson1972}}
	
\end{table*}

\clearpage

\section{Estimated values of the large separation, $\Delta\nu$, and their corresponding radial modes with $n$ = 1 to $n$ = 8, for each of the stars in the Kepler and TESS samples.}
\label{AppendixB}

\begin{table*}[h!]
\centering
    \caption{Estimated $\Delta\nu$ and the corresponding frequency ranges for the radial modes with $n=1$ to $n=8$, for the Kepler sample.}
    \label{tb:seismic_parameters_Kepler}
    \renewcommand{\arraystretch}{1.5}
	\addtolength{\tabcolsep}{2.5 pt}
	\resizebox{\textwidth}{!}{
    \begin{tabular}{cccccccccc}
    \hline
         KIC & $\Delta\nu$ (d$^{-1}$) & $f_{1}$(d$^{-1}$) & $f_{2}$(d$^{-1}$) & $f_{3}$(d$^{-1}$) & $f_{4}$(d$^{-1}$) & $f_{5}$(d$^{-1}$) & $f_{6}$(d$^{-1}$) & $f_{7}$(d$^{-1}$) & $f_{8}$(d$^{-1}$)\\
    \hline
         3219256 & 5.64$\pm$0.09 & [17.153, 20.408] & [22.192, 25.66] & [27.249, 30.96] & [32.239, 36.115] & [37.479, 41.613] & [42.987, 47.259] & [48.553, 52.945] & [54.167, 58.591]\\
         3760002 & 3.29$\pm$0.09 & [10.389, 12.165] & [13.086, 15.271] & [16.170, 18.733] & [19.454, 22.248] & [22.702, 25.579] & [25.840, 28.965] & [28.911, 32.303] & [32.303, 35.765]\\
         3942911  & 5.12$\pm$0.09 & [15.621, 18.809] & [20.249, 23.645] & [24.742, 28.653] & [29.362, 33.428] & [34.155, 38.575] & [39.291, 43.676] & [44.189, 48.757] & [49.200, 53.47]\\
         5428254 & 5.12$\pm$0.09 & [15.689, 18.818] & [20.217, 23.674] & [24.992, 28.553] & [29.300, 33.515] & [34.120, 38.501] & [39.149, 43.637] & [44.489, 48.681] & [49.197, 53.661]\\
         5474427 & 2.60$\pm$0.09 & [7.772, 9.389] & [10.055, 11.834] & [12.566, 14.587] & [15.189, 17.338] & [17.711, 19.958] & [20.297, 22.618] & [22.719, 25.269] & [25.269, 28.069]\\
         5774557 & 2.86$\pm$0.09 & [8.652, 10.388] & [11.124, 13.219] & [13.953, 16.195] & [16.739, 19.174] & [19.581, 22.209] & [22.413, 25.076] & [25.032, 27.96] & [27.960, 29.909]\\
         7106205 & 3.38$\pm$0.17 & [9.982, 12.698] & [13.325, 16.053] & [16.453, 19.462] & [19.462, 22.450] & [22.450, 26.79] & [25.797, 26.790] & - & - \\
         7217483 & 2.86$\pm$0.09 & [8.557, 10.385] & [11.094, 13.201] & [13.900, 16.119] & [16.707, 19.114] & [19.678, 22.075] & [22.287, 25.078] & [24.963, 27.861] & [27.861, 30.917]\\
         7548479 & 5.20$\pm$0.09 & [16.119, 18.799] & [20.555, 23.653] & [25.381, 28.906] & [30.135, 33.762] & [35.110, 38.578] & [40.069, 44.198] & [45.160, 49.432] & [50.187, 54.41]\\
         8103917 & 3.64$\pm$0.09 & [11.379 11.381] & [15.173, 16.997] & [18.004, 20.757] & [21.584, 24.486] & [26.0150, 28.062] & [28.723, 31.500] & [32.038, 35.603] & [35.981, 39.442]\\
         8446738 & 5.12$\pm$0.17 & [15.770, 15.772] & [20.415, 20.417] & [24.560, 29.002] & [29.002, 33.554] & [34.023, 38.951] & [38.726, 43.748] & [43.748, 48.649] & [48.649, 54.12]\\
         8827821 & 5.29$\pm$0.09 & [16.090, 19.139] & [20.862, 24.383] & [25.590, 29.461] & [30.251, 34.312] & [35.175, 39.457] & [40.411, 44.838] & [45.784, 50.058] & [51.558, 55.259]\\
         9489590 & 5.03$\pm$0.09 & [15.493, 18.461] & [20.056, 23.338] & [24.443, 28.285] & [29.228, 32.864] & [33.92, 38.084] & [38.955, 42.841] & [44.071, 47.508] & [48.572, 52.89]\\
         9762713 & 3.38$\pm$0.09 & [10.567, 12.521] & [13.585, 15.261] & [16.661, 19.174] &[20.097, 22.799] & [23.439, 26.043] & [26.758, 29.151] & - & [34.262, 34.264]\\
         10549371 & 3.64$\pm$0.09 & [11.360, 13.523] & [14.496, 17.079] & [18.307, 20.729] & [22.103, 24.592] & [25.163, 27.759] & - & [33.978, 35.634] & - \\
         10686752 & 5.81$\pm$0.09 & [18.038, 20.221] & [24.321, 26.206] & [28.316, 31.707] & [33.049, 36.194] & [39.092, 42.408] & [44.279, 48.556] & [50.190, 53.95] & [56.651, 60.104]\\
         11183539 & 4.07$\pm$0.09 & [12.524, 15.240] & [16.153, 19.159] & [19.896, 23.261] & [23.717, 27.35] & [27.633, 31.519] & [31.559, 35.643] & [35.523, 39.698] & [39.514, 43.937]\\
         12647070 & 4.07$\pm$0.17 & [12.350, 15.126] & [15.850, 19.51] & [19.691, 23.378] & [23.378, 27.561] & [27.166, 32.227] & [31.165, 35.393] & [35.393, 39.984] & [39.168, 43.152]\\
         \end{tabular}
         }

\end{table*}

\begin{table*}[h!]
\centering
    \caption{Same as Table~\ref{tb:seismic_parameters_Kepler} for the TESS\ sample.}
    \label{tb:seismic_parameters_TESS}
    \renewcommand{\arraystretch}{1.5}
	\addtolength{\tabcolsep}{2.5 pt}
	\resizebox{\textwidth}{!}{
    \begin{tabular}{cccccccccc}
    \hline
         TIC & $\Delta\nu$ (d$^{-1}$) & $f_{1}$(d$^{-1}$) & $f_{2}$(d$^{-1}$) & $f_{3}$(d$^{-1}$) & $f_{4}$(d$^{-1}$) & $f_{5}$(d$^{-1}$) & $f_{6}$(d$^{-1}$) & $f_{7}$(d$^{-1}$) & $f_{8}$(d$^{-1}$)\\
    \hline
         2096047 & 3.89$\pm$0.26 & [11.621, 14.920] & [14.794, 18.871] & [18.548, 22.693] & [21.952, 25.772] & [25.772, 29.661] & [29.660, 29.662] & - & -\\
         3697787 & 5.36$\pm$0.17 & [16.453, 19.355] & [21.253, 24.341] & [25.849, 28.017] & - & [37.760, 37.762] & - & - & -\\
         37908913  & 5.10$\pm$0.26 & [15.443, 18.108] & [20.367, 24.038] & [24.353, 28.758] & [28.758, 33.355] & [33.355, 36.897] & - & [45.252, 45.881] & -\\
         69025963 & 5.96$\pm$0.35 & [20.710, 20.712] & [23.027, 26.849] & [29.596, 32.305] & [36.185, 37.648] & [39.733, 44.392] & [44.391, 44.393] & - & -\\
         86893888 & 5.79$\pm$0.26 & [17.412, 21.113] & [23.888, 26.662] & [27.424, 31.252] & [36.955, 36.957] & [39.761, 39.763] & - & - & -\\
         121597646 & 3.54$\pm$0.26 & [10.779, 13.641] & [13.411, 17.265] & [16.798, 21.283] & [20.020, 24.673] & [23.354, 28.794] & [26.573, 32.417] & [29.745, 35.813] & [33.609, 36.609]\\
         122069152 & 4.41$\pm$0.26 & [13.765, 17.011] & [17.011, 21.458] & [21.458, 26.146] & [25.109, 30.105] & [29.25, 34.383] & [33.226, 39.618] & [39.618, 42.401] & [41.800, 42.401] \\
         155457396 & 3.89$\pm$0.17 & [11.980, 14.899] & [15.249, 18.297] & [19.129, 22.917] & [22.578, 26.829] & [26.112, 31.056] & [29.845, 34.852] & [33.635, 34.852] & -\\
         158787200 & 3.02$\pm$0.17 & [9.178, 11.470] & [11.697, 14.334] & [15.466, 17.629] & [17.495, 21.024] & [20.265, 23.223] & [23.145, 26.425] & [26.425, 29.719] & [29.718, 29.720]\\
         231148059 & 3.37$\pm$0.17 & [10.444, 12.344] & [14.481, 15.184] & [17.295, 18.991] & [19.871, 22.746] & [22.746, 25.471] & [27.815, 28.699] & - & -\\
         232064019 & 4.15$\pm$0.17 & [12.753, 15.781] & [16.425, 19.939] & [20.485, 24.218] & [24.218, 28.168] & [28.168, 32.143] & [32.143, 36.731] & [36.674, 41.323] & [41.322, 41.324]\\
         255733180 & 2.76$\pm$0.26 & [7.882, 10.210] & [10.099, 13.640] & [12.566, 16.716] & [15.187, 19.753] & [17.703, 22.826] & [20.189, 25.834] & [22.674, 28.747] & [25.257, 31.493]\\
         316826948 & 5.62$\pm$0.26 & [17.467, 20.497] & [21.779, 25.765] & [26.489, 31.108] & [31.876, 34.748] & [37.135, 42.488] & [42.488, 45.544] & [51.372, 51.374] & -\\
         387226824 & 5.36$\pm$0.17 & [16.591, 18.981] & [21.007, 24.738] & [25.884, 29.657] &[30.694, 34.456] & [35.430, 39.207] & [41.528, 43.106] & [47.375, 47.377] & -\\
         394015973 & 3.46$\pm$0.17 & [10.336, 13.155] & [13.411, 16.621] & [16.726, 20.298] & [20.064, 23.963] & [23.258, 27.678] & [26.511, 31.234] & [29.757, 34.853] & [33.209, 38.519] \\
         397012320 & 4.75$\pm$0.17 & [14.152, 18.159] & [18.844, 22.899] & [22.516, 27.513] & [26.877, 31.867] & [31.296, 35.838] & [35.837, 35.839] & - & [49.141, 49.143]\\
         398733851 & 3.28$\pm$0.17 & [9.915, 12.82] & [12.288, 15.485] & [15.485, 19.656] & [18.634, 23.379] & [21.453, 26.849] & [24.471, 30.413] & [28.129, 33.936] & [30.490, 36.873]\\
         
         \end{tabular}
         }

\end{table*}

\clearpage

\section{Parameters of the modelled stars}
\label{AppendixC}
\begin{table*}[h!]
    
	\caption{Parameters of the seismically constrained models corresponding to the Kepler sample, with their associated $1\sigma$ uncertainties.}
    \renewcommand{\arraystretch}{1.5}
	\addtolength{\tabcolsep}{2.5 pt}
	\resizebox{\textwidth}{!}{
    \begin{tabular}{cccccccccc}
			\hline
			KIC & Z & $\mathrm{log\ (L/L_{\odot})}$ & $\mathrm{M\ (M_{\odot})}$ & $\mathrm{R\ (R_{\odot})}$ & $\bar \rho\ (\bar \rho_{\odot})$ & \logg & $\mathrm{T_{eff}}$ (K) & v (\kms) & Age (\myr) \\
			\hline
	       	3219256 & $0.021\pm0.005$ & $0.91\pm0.03$ & $1.66\pm0.06$ & $1.69\pm0.05$ & $0.34\pm0.02$ & $4.20\pm0.02$ & $7505\pm85$ & $122\pm67$ & $436\pm265$ \\
			3760002 & $0.020\pm0.006$ & $1.17\pm0.05$ & $1.84\pm0.10$ & $2.46\pm0.08$ & $0.12\pm0.01$ & $3.92\pm0.02$ & $7243\pm147$ & $111\pm58$ & $906\pm244$ \\
            3942911 & $0.021\pm0.006$ & $0.97\pm0.05$ & $1.70\pm0.08$ & $1.80\pm0.06$ & $0.29\pm0.02$ & $4.16\pm0.02$ & $7530\pm149$ & $132\pm69$ & $544\pm294$ \\
            5428254 & $0.021\pm0.006$ & $0.99\pm0.05$ & $1.72\pm0.08$ & $1.81\pm0.06$ & $0.29\pm0.02$ & $4.16\pm0.02$ & $7607\pm152$ & $133\pm69$ & $528\pm282$ \\
            5474427 & $0.022\pm0.005$ & $1.34\pm0.04$ & $2.02\pm0.08$ & $2.99\pm0.09$ & $0.08\pm0.01$ & $3.79\pm0.02$ & $7207\pm127$ & $101\pm51$  & $831\pm140$\\
            5774557 & $0.021\pm0.006$ & $1.28\pm0.05$ & $1.95\pm0.09$ & $2.76\pm0.09$ & $0.09\pm0.01$ & $3.85\pm0.02$ & $7266\pm147$ & $106\pm53$ & $858\pm185$ \\
            7106205 & $0.020\pm0.006$ & $1.06\pm0.04$ & $1.73\pm0.08$ & $2.36\pm0.09$ & $0.13\pm0.01$ & $3.93\pm0.03$ & $6898\pm79$ & $110\pm58$ & $1095\pm246$ \\
            7217483 & $0.021\pm0.006$ & $1.25\pm0.05$ & $1.92\pm0.09$ & $2.74\pm0.09$ & $0.09\pm0.01$ & $3.84\pm0.02$ & $7157\pm145$ & $105\pm53$ & $907\pm192$ \\
            7548479 & $0.021\pm0.006$ & $0.96\pm0.03$ & $1.69\pm0.07$ & $1.78\pm0.05$ & $0.30\pm0.02$ & $4.17\pm0.02$ & $7505\pm86$ & $132\pm69$ & $527\pm311$ \\
            8103917 & $0.020\pm0.006$ & $1.13\pm0.05$ & $1.82\pm0.09$ & $2.28\pm0.08$ & $0.15\pm0.01$ & $3.98\pm0.02$ & $7344\pm149$ & $120\pm63$ & $852\pm254$ \\
            8446738 & $0.022\pm0.006$ & $0.93\pm0.05$ & $1.67\pm0.06$ & $1.79\pm0.06$ & $0.29\pm0.02$ & $4.16\pm0.02$ & $7376\pm148$ & $133\pm69$ & $558\pm295$ \\
            8827821 & $0.021\pm0.006$ & $0.97\pm0.05$ & $1.71\pm0.08$ & $1.77\pm0.05$ & $0.31\pm0.02$ & $4.18\pm0.02$ & $7611\pm151$ & $136\pm69$ & $468\pm300$ \\
            9489590 & $0.022\pm0.006$ & $0.94\pm0.05$ & $1.68\pm0.07$ & $1.81\pm0.06$ & $0.29\pm0.02$ & $4.15\pm0.02$ & $7387\pm149$ & $133\pm69$ & $575\pm306$ \\
            9762713 & $0.020\pm0.006$ & $1.14\pm0.04$ & $1.81\pm0.09$ & $2.40\pm0.08$ & $0.13\pm0.01$ & $3.94\pm0.01$ & $7181\pm121$ & $112\pm59$ & $938\pm253$ \\
            10549371 & $0.021\pm0.006$ & $1.09\pm0.03$ & $1.78\pm0.08$ & $2.26\pm0.07$ & $0.15\pm0.01$ & $3.98\pm0.02$ & $7199\pm46$ & $120\pm62$ & $913\pm257$ \\
            10686752 & $0.021\pm0.006$ & $0.90\pm0.04$ & $1.65\pm0.06$ & $1.66\pm0.05$ & $0.36\pm0.02$ & $4.21\pm0.02$ & $7513\pm150$ & $111\pm61$ & $404\pm236$ \\
            11183539 & $0.020\pm0.006$ & $1.16\pm0.05$ & $1.86\pm0.09$ & $2.14\pm0.07$ & $0.19\pm0.02$ & $4.05\pm0.02$ & $7683\pm156$ & $128\pm66$ & $675\pm222$ \\
            12647070 & $0.020\pm0.006$ & $1.10\pm0.05$ & $1.80\pm0.09$ & $2.11\pm0.08$ & $0.19\pm0.02$ & $4.04\pm0.03$ & $7488\pm151$ & $127\pm65$ & $760\pm215$ \\

    \end{tabular}
    }
    \tablefoot{From left to right: KIC identifier, metallicity [Fe/H], luminosity (log(L/L$_\odot$)), mass (M/M$_\odot$), radius (R/R$_\odot$), mean density ($\bar{\rho}/\bar{\rho}\odot$), surface gravity (\logg), effective temperature (T$_\mathrm{eff}$), rotational velocity (v$_{\mathrm{rot}}$), and age (Myr).}
	\label{tb:Parameters_models_Kepler}
\end{table*}

\begin{table*}[h!]
    
    \centering
    
	\caption{Same as Table~\ref{tb:Parameters_models_Kepler} for the TESS\ sample.}
    \renewcommand{\arraystretch}{1.5}
	\addtolength{\tabcolsep}{2.5 pt}
	\resizebox{\textwidth}{!}{
    \begin{tabular}{cccccccccc}
			\hline
			TIC & Z & $\mathrm{log\ (L/L_{\odot})}$ & $\mathrm{M\ (M_{\odot})}$ & $\mathrm{R\ (R_{\odot})}$ & $\bar \rho\ (\bar \rho_{\odot})$ & \logg & $\mathrm{T_{eff}}$ (K) & v (\kms) & Age (\myr) \\
			\hline
	       	2096047 & $0.020\pm0.006$ & $1.05\pm0.05$ & $1.74\pm0.08$ & $2.15\pm0.09$ & $0.18\pm0.02$ & $4.02\pm0.03$ & $7218\pm82$ & $124\pm64$ & $895\pm223$ \\
			3697787 & $0.022\pm0.005$ & $0.91\pm0.03$ & $1.66\pm0.05$ & $1.74\pm0.05$ & $0.32\pm0.02$ & $4.18\pm0.02$ & $7397\pm62$ & $127\pm68$ & $486\pm292$ \\
            37908913 & $0.021\pm0.006$ & $0.95\pm0.03$ & $1.68\pm0.06$ & $1.79\pm0.06$ & $0.29\pm0.03$ & $4.16\pm0.02$ & $7441\pm64$ & $133\pm68$ & $562\pm297$ \\
            69025963 & $0.021\pm0.005$ & $0.88\pm0.03$ & $1.64\pm0.05$ & $1.64\pm0.05$ & $0.37\pm0.03$ & $4.22\pm0.02$ & $7499\pm60$ & $105\pm59$ & $331\pm240$ \\
            86893888 & $0.022\pm0.005$ & $0.87\pm0.03$ & $1.64\pm0.05$ & $1.66\pm0.05$ & $0.36\pm0.03$ & $4.21\pm0.02$ & $7404\pm58$ & $109\pm61$  & $387\pm240$\\
            121597646 & $0.020\pm0.006$ & $1.14\pm0.05$ & $1.82\pm0.09$ & $2.32\pm0.11$ & $0.15\pm0.02$ & $3.97\pm0.03$ & $7322\pm78$ & $118\pm62$ & $865\pm251$ \\
            122069152 & $0.020\pm0.006$ & $1.04\pm0.04$ & $1.75\pm0.08$ & $1.99\pm0.08$ & $0.22\pm0.02$ & $4.08\pm0.03$ & $7471\pm84$ & $129\pm66$ & $728\pm236$ \\
            155457396 & $0.020\pm0.006$ & $1.09\pm0.04$ & $1.78\pm0.08$ & $2.16\pm0.08$ & $0.18\pm0.02$ & $4.02\pm0.03$ & $7359\pm73$ & $125\pm64$ & $834\pm236$ \\
            158787200 & $0.020\pm0.006$ & $1.24\pm0.04$ & $1.91\pm0.09$ & $2.63\pm0.10$ & $0.11\pm0.01$ & $3.88\pm0.03$ & $7286\pm84$ & $107\pm56$ & $859\pm211$ \\
            231148059 & $0.020\pm0.006$ & $1.19\pm0.04$ & $1.86\pm0.09$ & $2.42\pm0.09$ & $0.13\pm0.01$ & $3.94\pm0.03$ & $7350\pm87$ & $114\pm60$ & $854\pm243$ \\
            232064019 & $0.022\pm0.005$ & $0.95\pm0.04$ & $1.66\pm0.06$ & $2.03\pm0.07$ & $0.20\pm0.02$ & $4.05\pm0.03$ & $7003\pm62$ & $124\pm64$ & $951\pm238$ \\
            255733180 & $0.021\pm0.006$ & $1.32\pm0.05$ & $2.00\pm0.10$ & $2.84\pm0.15$ & $0.09\pm0.01$ & $3.83\pm0.03$ & $7328\pm78$ & $105\pm53$ & $805\pm168$ \\
            316826948 & $0.021\pm0.005$ & $0.91\pm0.03$ & $1.66\pm0.06$ & $1.69\pm0.05$ & $0.34\pm0.03$ & $4.20\pm0.02$ & $7477\pm60$ & $118\pm65$ & $438\pm251$ \\
            387226824 & $0.021\pm0.005$ & $0.92\pm0.03$ & $1.67\pm0.06$ & $1.74\pm0.05$ & $0.32\pm0.02$ & $4.18\pm0.02$ & $7443\pm58$ & $131\pm68$ & $464\pm297$ \\
            394015973 & $0.020\pm0.006$ & $1.08\pm0.04$ & $1.75\pm0.08$ & $2.33\pm0.09$ & $0.14\pm0.01$ & $3.95\pm0.03$ & $7027\pm65$ & $114\pm60$ & $1012\pm250$ \\
            397012320 & $0.022\pm0.006$ & $0.95\pm0.04$ & $1.67\pm0.06$ & $1.87\pm0.07$ & $0.26\pm0.03$ & $4.12\pm0.03$ & $7284\pm75$ & $131\pm67$ & $693\pm280$ \\
            398733851 & $0.020\pm0.006$ & $1.10\pm0.05$ & $1.77\pm0.09$ & $2.42\pm0.12$ & $0.13\pm0.02$ & $3.92\pm0.03$ & $6997\pm75$ & $109\pm57$ & $1029\pm241$ \\
                   
    \end{tabular}
    }
    
	\label{tb:Parameters_models_TESS}
\end{table*}

\label{AppendixD}

\end{appendices}

\end{document}